\documentclass[iop]{emulateapj}
\usepackage[version=3]{mhchem}
\usepackage{longtable}
\usepackage{apjfonts,natbib}

\makeatletter
\newcounter{reaction}
\renewcommand\thereaction{C\,\arabic{reaction}}
\newcommand\reactiontag{\refstepcounter{reaction}\tag{\thereaction}}
\newcommand\reaction@[2][]{\begin{equation}\ce{#2}%
\ifx\@empty#1\@empty\else\label{#1}\fi%
\reactiontag\end{equation}}
\newcommand\reaction@nonumber[1]{\begin{equation*}\ce{#1}%
\end{equation*}}
\newcommand\reaction{\@ifstar{\reaction@nonumber}{\reaction@}}
\makeatother

\shorttitle{Reflected Light Spectra of Cold Giant Exoplanets}

\shortauthors{Hu}

\begin{document}

\title{Information in the Reflected Light Spectra of Widely Separated Giant Exoplanets}

\author{Renyu Hu$^{1,2}$}
\affil{$^1$Jet Propulsion Laboratory, California Institute of Technology, Pasadena, CA 91109}
\affil{$^2$Division of Geological and Planetary Sciences, California Institute of Technology, Pasadena, CA 91125}
\email{renyu.hu@jpl.nasa.gov \\ Copyright 2019. All rights reserved.}

\begin{abstract}
Giant exoplanets located $>1$ AU away from their parent stars have atmospheric environments cold enough for water and/or ammonia clouds. We have developed a new equilibrium cloud and reflected light spectrum model, ExoREL, for widely separated giant exoplanets. The model includes the dissolution of ammonia in liquid water cloud droplets, an effect studied for the first time for exoplanets. While preserving the causal relationship between temperature and cloud condensation, ExoREL is simple and fast to enable efficient exploration of parameter space. Using the model, we find that the mixing ratio of methane and the cloud top pressure of a giant exoplanet can be uniquely determined from a single observation of its reflected light spectrum at wavelengths less than 1 $\mu$m if it has a cloud deck deeper than $\sim0.3$ bars. This measurement is enabled by the weak and strong bands of methane and requires a signal-to-noise ratio of 20. The cloud pressure once derived, provides information about the internal heat flux of the planet. Importantly, we find that for a low, Uranus-like internal heat flux, the planet can have a deep liquid water cloud, which will sequester ammonia and prevent the formation of the ammonia cloud that would otherwise be the uppermost cloud layer. This newly identified phenomenon causes a strong sensitivity of the cloud top pressure on the internal heat flux. Reflected light spectroscopy from future direct-imaging missions therefore not only measure the atmospheric abundances but also characterize the thermal evolution of giant exoplanets.
\end{abstract}

\keywords{ Extrasolar gas giants --- Extrasolar ice giants --- Exoplanet atmospheric composition --- Direct imaging --- Exoplanet evolution}

\section{Introduction}

The discovery of more than a thousand exoplanets has greatly extended the horizon of planetary exploration \citep[e.g.,][]{Howard2013, Batalha2014, Marcy2014}. The transit spectra of short-period giant exoplanets and several Neptune- and sub-Neptune-sized exoplanets orbiting low-mass stars have been observed. The spectra reveal the thermal emission of the planets or the transmission through their atmospheres \citep[e.g.,][]{Seager2010b}. These measurements have indicated molecular absorptions of \ce{H2O}, \ce{CO}, \ce{CH4}, and \ce{CO2}, and in some cases, the effects of clouds and hazes in the atmospheres \citep[e.g., see the review of][and references therein]{Burrows2014}. The current observations of exoplanet atmospheres using the transit technique work best for planets close to their parent stars. Due to stellar irradiation, these planets generally have warm and hot atmospheres that are very different from any planetary atmospheres in the Solar System \citep{Burrows1997, Seager1998}.

Future direct-imaging exoplanet space missions will provide the capability to directly detect exoplanets of nearby stars. WFIRST, a 2.4-m space telescope being developed, will be equipped with an internal coronagraph that can image giant planets around nearby stars \citep[e.g.,][]{Douglas2018spie}. Flying a starshade in formation with WFIRST as an external occulter to suppress starlight will enable imaging Earth-sized planets, and also obtaining reflected light spectra of a handful of known giant planets \citep{Seager2018SRM}. Two of the four flagship mission concepts that are being considered for the 2020 Astrophysics Decadal Survey, HabEx and LUVOIR, plan to directly image exoplanets and measure their spectra. The common feature of WFIRST and exoplanet direct imaging mission concepts is that they would obtain reflected light spectra of exoplanets at visible and near-infrared wavelengths. The inner working angles -- the smallest angle at which a planet can be detected -- determine that the exoplanets to be observed are sufficiently separated from their parent stars. These exoplanets will thus have atmospheres much colder and different chemical states than most of the atmospheres observed currently with the transit technique. 

A great diversity of the possible spectral features in the reflected light of exoplanets can be anticipated as a result of clouds and gases in their atmospheres. Rayleigh scattering, molecular absorption, and scattering and absorption by atmospheric condensates determine the reflection spectra of gaseous exoplanets \citep{Marley1999, Seager2000b}. Whether there exist clouds is the primary factor that controls the appearance of an exoplanet. Depending on the atmospheric temperature, an exoplanet may or may not have clouds. Assuming an atmospheric elemental abundance the same as the Sun, giant exoplanets may have ammonia, water, or silicate clouds in their atmospheres depending on the orbital distances from their parent stars \citep{Marley1999,Sudarsky2000,Sudarsky2003,Burrows2004}. The radiative properties of the clouds are sensitive to the vertical extent of the cloudy layer and the sizes of cloud particles \citep{Ackerman2001}. The elemental abundance of the atmosphere also affects the formation of the clouds and the spectra \citep{Cahoy2010}. As such, reflected light spectra of exoplanets contain rich information on the composition and energetic and dynamic processes of exoplanet atmospheres.

The reflected spectra measured by first-generation direct imaging space missions will probably be similar to those spectra obtained for Solar System giant planets in the 1970s. The spectra between 600 and 1000 nm contain information on the compositions and cloud structures in the atmospheres of giant planets in the Solar System. For example, the spectrum of Jupiter that contains strong, intermediate, and weak methane absorption bands has been used to reject simple models of a single reflective cloud deck but indicate a more complex double-layer cloud structure \citep{Sato1979}. Comparing the spectrum of the center and that of the limb further determines the vertical extent of the upper cloud layer \citep{Sato1979}. With the methane mixing ratio known from the ratio of the strengths between the \ce{H2} quadruple lines and the methane absorption bands, characterization of the cloud structure on Jupiter is also possible at a rather low spectral resolution of $\sim30$. \citet{Banfield1998} uses narrow-band images of Jupiter obtained by the Galileo spacecraft to constrain that the upper cloud layer is at $750\pm200$ mbar, and that a haze layer exists above the upper cloud layer (i.e., the upper tropospheric haze). The optical depth of this upper cloud layer is highly varied by location, ranging from 0 to more than 20 \citep{Banfield1998, Matcheva2005}. The composition of the upper cloud layer is inferred to be ammonia, consistent with the prediction of equilibrium cloud models \citep{Weidenschilling1973,Atreya1999}. 

The rich history of Solar System investigations shows that a combination of intermediate resolution spectra, radiative-transfer spectral analysis, and forward modeling of atmospheric chemistry and cloud physics can lead to important insight into the atmospheres on gaseous planets that include Jupiter- and Neptune-sized exoplanets. To prepare for the future exoplanet direct imaging observations, we are motivated to use a hierarchy of radiative transfer and atmospheric chemistry models to address one of the key questions: what could we learn about the planets from their reflected light spectra at a modest spectral resolution? In this paper, we focus on the measurement of cold giant exoplanet's atmospheric compositions as they are among the first targets of direct imaging \citep[e.g.,][]{Spergel2013}. We do not consider the effects of haze in this paper, as it is intricate to model with high fidelities and the issue of haze should be studied in separate papers \citep[e.g.,][]{gao2017sulfur}. 

Several models exist for cold giant exoplanets, and the state of the art has been summarized in a set of reports commissioned by the NASA Exoplanet Exploration Program in 2014 to support direct imaging missions \citep{Burrows2014report,Hu2014report,Marley2014report}. Early investigations of reflection spectra of extrasolar giant planets have found a great diversity in the possible spectral features as a result of the competition of cloud and gas opacities. \citet{Marley1999} first calculated the reflection spectra of extrasolar giant planets using the atmosphere models of \citet{Burrows1997}, which did not include the deposition of stellar radiation. After that, more self-consistent atmospheric models with temperature profiles determined by the irradiation and the composition have been applied to the study of exoplanets. \citet{Seager2000b} presented the reflection spectra of close-in extrasolar gas giants (i.e., hot Jupiters) with the treatment of silicate clouds. \citet{Sudarsky2000, Sudarsky2003, Burrows2004} further explored giant exoplanets with varied orbital distance from their parent stars and classified the planets according to the existence of ammonia, water, or silicate clouds. They assumed gas abundances of solar metallicity at thermochemical equilibrium and employed a simple cloud model. \citet{Cahoy2010} improved the work by simulating cloud microphysics based on the method of \citet{Ackerman2001} that calculates the height, particle sizes, and consequently optical thickness of water and ammonia clouds, for atmospheres with super-solar metallicities (up to 30 times solar metallicities). Using a similar model, \citet{MacDonald2018} argued that absorption features of water would manifest prominently in the reflected light spectra of some planets that are hotter than Jupiter.   

There is a need for a fast and self-consistent model to calculate the atmospheric structure and reflected light spectra for cold exoplanets. The full non-grey radiative-convective calculation that couples with a cloud microphysics calculation takes substantial computational time to converge and it is less suitable to explore wide ranges of semi-major axes, atmospheric metallicities, and internal heat flux. For instance, \citet{Cahoy2010} only presented a handful of cases, and \citet{MacDonald2018} explored the reflected light spectra across a wide parameter space but had to circumvent the computation of self-consistent temperature profiles by a parametric model fit to a few tens of previously calculated profiles. The need for a simple model is perhaps more evident when it comes to atmospheric retrieval. Several retrieval frameworks have been published for giant planets' reflected light spectra \citep{Lupu2016,Nayak2017,Lacy2019}. While these results are quite promising, they uniformly had to decouple the atmospheric temperature and the cloud density -- the physical quantities that are intimately tied to each other -- to achieve a computationally viable retrieval. We are therefore motivated the provide a fast and self-consistent model, ExoREL, in which we preserve the causal relationship between condensation and clouds, and we simplify the calculations on the pressure-temperature profiles and cloud microphysics. In terms of the hierarchy of complexities, ExoREL is a physically plausible, and yet minimally complex model.

Another innovative aspect of our model is the treatment of dissolution of \ce{NH3} is water droplets. The giant planets in the Solar System have clouds made of aqueous \ce{NH3} solutions. On Jupiter and Saturn, the bottom part of the water cloud is predicted to have the liquid form that dissolves \ce{NH3}, if the atmospheric metallicity is greater than 3 -- 5 times the solar metallicity \citep[e.g.][]{Weidenschilling1973,Atreya1999}. The solubility of \ce{NH3} in liquid water clouds is the leading hypothesis to explain the observed depletion of \ce{NH3} in the upper atmospheres of Uranus and Neptune \citep{DePater1989,Romani1989}. To our knowledge, this effect has not been included in exoplanet studies. This effect would be relevant for cool exoplanets in which water clouds can exist, corresponding to the equilibrium temperature less than approximately 320 K. For planets of FGK stars, this corresponds to wide orbital separations (e.g., $>0.7$ AU for Sun-like stars). The paper is organized as follows. We describe ExoREL in \S~2 and the results in \S~3. We discuss the implications of the results in \S~4 and conclude in \S~5.

\section{Model}

We have developed a simplified calculation of the cloud top pressure on gaseous exoplanets that have H$_2$-dominated atmospheres and have equilibrium temperatures between 100 and 300 K. The model, called ExoREL (Exoplanet REflected Light), is an extension of the classical equilibrium cloud model that has successfully predicted the bulk cloud structure of Jupiter \citep{Weidenschilling1973, Atreya1999}. The model considers water and ammonia as potential condensable species, estimates the particle size to calculate the radiative properties of clouds, includes the cloud feedback on the adiabatic lapse rate and the planetary albedo, and computes disk-averaged reflected light spectra at any phase angle of observation. We have validated the model by reproducing the temperature structure, the upper cloud structure, and the disk-averaged reflection spectrum of Jupiter (see \S~\ref{sec:validation}). The current model does not include \ce{NH4SH} as a possible condensable gas, even though a \ce{NH4SH} cloud has been predicted to exist in the atmosphere of Jupiter and probably accounts for the lack of \ce{H2S} features in disk-integrated spectra of Jupiter in the near-infrared \citep[e.g.,][]{Lewis1969,Weidenschilling1973,atreya1985photochemistry,Atreya1999,Atreya2003,Wong2004}.

We assume water, methane, and ammonia are always the dominant carrier for oxygen, carbon, and nitrogen. This assumption is valid for the planets of consideration (i.e., cold Jupiter and Neptune-sized planets having atmospheres mainly composed of hydrogen and helium). The hydrogen dominance and low temperature of the atmosphere ensure these elements in their most hydrogenated forms \citep{Madhusudhan2012,Hu2014}. This way, we do not need to solve the full thermochemical equilibrium balance. Note that this assumption breaks down in two scenarios. One is that the metallicity in the atmosphere cannot be too high; otherwise the atmosphere would be in the hydrogen-poor regime in which water, ammonia, and methane are no longer the dominant trace gases \citep{Moses2013, Hu2014}. The other is that the planet cannot have very large internal heat flux, otherwise the temperature would be too high at the quenching pressure of water, methane, and ammonia \citep{Hu2014}. All models in this paper apply to matured \ce{H2}-dominated giant exoplanets at wide orbital separations from their parent stars (e.g., 1 --10 AU of Sun-like stars).

\subsection{Pressure-Temperature Profile} \label{sec:tp}

We calculate the pressure-temperature profile of the atmosphere using the grey-atmosphere formulation derived by \cite{Guillot2010}. The grey-atmosphere approximation, if parameterized appropriately, can lead to fast calculations of the pressure-temperature profile that is close to the result from non-grey radiative-convective calculations for irradiated gas giants \citep{Fortney2008,hansen2008absorption,Guillot2010,Parmentier2015}, especially in terms of the radiative-convective boundary and the adiabatic portion of the pressure-temperature profile beneath that boundary. Some of the more complex grey-atmosphere models can produce temperature inversion in the upper atmosphere \citep{Parmentier2015}. Here we adopt the simple model of \cite{Guillot2010}, because it allows fast exploration of controlling parameters, and it easily incorporates the effect of changing atmospheric metallicities. The grey-atmosphere model adopted in this work does not produce a temperature inversion, but this is less of a concern when it comes to interpreting reflected light spectra because the spectra are insensitive to a temperature inversion. The planets of interest are not tidally synchronized, so we assume full heat redistribution for the calculation of temperature profiles.

The input parameters of the temperature-pressure profile calculation are the gravitational acceleration $g$, the metallicity $[M/H]$, the irradiation temperature $T_{\rm irr}$, and the intrinsic temperature $T_{\rm int}$. The latter two parameters describe the energy flux from stellar irradiation and internal heating, respectively. The optical depth of the atmosphere is calculated from the Rosseland mean opacity, taken from \citet{Freedman2014}. The grey-atmosphere model also employs an additional factor, $\gamma$, to describe the ratio between the opacity for stellar incident radiation and that for outgoing thermal radiation. This factor, therefore, controls the extent of greenhouse warming in the atmosphere. Following \cite{Guillot2010} we adopt a scaling relationship $\gamma\propto\sqrt{T_{\rm irr}}$, and assume $\gamma=0.07$ at $T_{\rm irr}\sim600$ K based on a strong greenhouse effect suggested for the exoplanet GJ~1214~b \citep{MillerRicci2010}.

\subsection{Adiabatic Lapse Rate and Cloud Density} \label{sec:cloudden}

We determine condensation of water vapor or ammonia by comparing their partial pressure to the saturation vapor pressure, and if condensation occurs, include the effects of condensation on the pressure-temperature profile by changing the adiabatic lapse rate.

The dry adiabatic lapse rate, $\Gamma_d$, is calculated from the heat capacity, which is gathered from the NIST Chemistry Webbook (http://webbook.nist.gov/chemistry/). We take into the account the temperature-dependent molar heat capacities of \ce{H2}, \ce{He}, \ce{H2O}, \ce{CH4}, and \ce{NH3}.

When condensation occurs, the moist adiabatic lapse rate ($\Gamma_m$) applies. The moist adiabatic lapse rate can be derived from the first law of thermodynamics, hydrostatic equilibrium, and the Clausius-Clapeyron equation. The result is
\begin{equation}
\Gamma_m = \Gamma_d \bigg(\frac{1+\frac{L_w \mu_wX_w}{RT}+\frac{L_a \mu_aX_a}{RT}}{1+\frac{L_w^2\mu_w^2X_w}{C_p\mu RT^2}+\frac{L_a^2\mu_a^2X_a}{C_p\mu RT^2}}\bigg),
\end{equation}
where $L$ is the latent heat of phase transition, $\mu_w$ and $\mu_a$ is the molar mass of water and ammonia, $X$ is the molecular mixing ratio, $C_p$ is the specific heat capacity of the atmosphere, $\mu$ is the mean molar mass of the atmosphere, $R$ is the universal gas constant, and $T$ is the atmosphere's temperature. We use subscripts $_w$ and $_a$ to denote the quantities for water and ammonia, respectively. This equation is valid for the diluted atmospheres in which the mixing ratios of condensable species are small. More terms must be included for the general expression of the moist adiabatic lapse rate \citep[e.g.,][]{li2018moist}. For simplicity the latent heat of water evaporation at 0 degree celsius is used for water phase transition above freezing and the latent heat of water sublimation at 0 degree celsius is used for water phase transition below freezing. This choice suffices because the latent heat only weakly depends on temperature. 

When a temperature profile is calculated by the grey-atmosphere formula, the profile is checked against convective instability and condensation. If either occurs, the temperature profile is modified to account for the dry or moist adiabatic lapse rates. If condensation occurs, we also calculate the cloud density similarly as \cite{Weidenschilling1973, Atreya1999}. We outline the main steps of this calculation as follows.
\begin{itemize}
\item From the highest pressure level ($10^9$ Pa) to the lowest pressure level (0.1 Pa), the partial pressures of water and ammonia, initially assumed to be the ones corresponding to the metallicity, are compared with their saturation vapor pressures. Their mixing ratios deep down in the atmosphere are set by the metallicity.
\item If none of the gases are saturated, the temperature gradient is compared with the dry adiabatic lapse rate. If the temperature gradient is greater than the dry adiabatic lapse rate, convection occurs and the temperature of the immediate layer above is adjusted according to the dry adiabatic lapse rate. The cloud density is zero and the mixing ratios of water and ammonia are unchanged.
\item If either or both of the gases are saturated, condensation occurs and the mixing ratio of the condensable species is reduced to the one corresponding to 100\% relative humidity ($X_w'$). The cloud density is then calculated assuming the masses of the ``missing'' water vapor (or ammonia) go to the condensed phase, and the formula is
\begin{equation}
\rho_c = \frac{(X_w-X_w')\mu_wP_i}{RT}.
\end{equation}
The temperature gradient is compared with the moist adiabatic lapse rate. If the temperature gradient is greater than the moist adiabatic lapse rate, convection occurs and the temperature of the immediate layer above is adjusted according to the moist adiabatic lapse rate. The mixing ratio of the condensed species in the immediate layer above is set to the saturation mixing ratio of this layer.
\item We calculate dissolution ammonia into water droplets. When water condenses to form droplets, the amount of ammonia at the layer is partitioned into the dissolved phase and the gas phase. The mass ratio between these two phases is determined by Henry's law \citep{Seinfeld2006}. We also calculate the dissociation of ammonia in the water droplet, which greatly enhances the nominal Henry's law constant, with the pH value of the cloud droplet self-consistently determined.
\item The steps above continue to the uppermost layer of the atmosphere. The result is a modified temperature profile according to appropriate lapse rates, mixing ratio profiles for water and ammonia that account for potential cold traps in the atmosphere, and cloud density profiles of water and ammonia the atmosphere.
\end{itemize}

As described above, we allow condensation to occur when a species reaches saturation (i.e., assuming condensation nuclei exist), and calculate the cloud density as if the condensed material stays in the atmospheric layer where condensation occurs. In cloud microphysics, this corresponds to the conditions that the sedimentation velocity of cloud particles matches the updraft velocity due to turbulent mixing. This condition also determines the cloud particle size (Section \ref{sec:particlesize}). \citet{Ackerman2001} introduced a tuning parameter ($f_{\rm rain}$) to choose the sedimentation velocity and the corresponding particle size. Our simple but self-consistent treatment here is comparable to theirs with $f_{\rm rain}\sim3$ for clouds at 0.1 -- 1 bar. Using Jupiter as a point of calibration, $f_{\rm rain}\sim3$ indeed produces the closest match to the vertical optical depth of the planet's ammonia cloud layer.

\subsection{Particle Sizes and Radiative Effects of Clouds} \label{sec:particlesize}

We estimate the sizes of cloud particles by the mass balance between sedimentation and updraft, following \citet{rossow1978cloud}. This also corresponds to the ``turbulent'' case in \cite{Marley1999}. Based on the sedimentation velocity of an ensemble of cloud particles following the lognormal distribution \citep{Seinfeld2006}, we have derived the following formula to estimate the quadratic mean particle diameter
\begin{eqnarray}
\overline{D_S} & =  &4.24 \bigg[\frac{\nu u }{\rho_{\rm p}gC_c}\bigg]^{1/2}, \nonumber \\
 & = & 0.95 C_c^{-1/2} \bigg(\frac{\nu}{10^{-5}\ {\rm Pa}\ {\rm s} }\bigg)^{1/2} \bigg(\frac{u}{10^{-4}\ {\rm m}\ {\rm s}^{-1} }\bigg)^{1/2}\nonumber\\
&&\times\bigg(\frac{\rho_{\rm p}}{2000\ {\rm kg}\ {\rm m}^{-3} }\bigg)^{-1/2}
\bigg(\frac{g}{10\ {\rm m}\ {\rm s}^{-2} }\bigg)^{-1/2} \ {\rm \mu m}, \label{eq:particlesize}
\end{eqnarray}
where $\nu$ is the viscosity of the atmosphere, $u$ is the updraft velocity, $\rho_{\rm p}$ is the density of the condensable material, $g$ is the gravitational acceleration, and $C_c$ is the slip parameter. See Appendix \ref{app:particlesize} for the derivation of this formula. The particle size has strong pressure dependency, mostly via the slip parameter $C_c$ that depends on the mean free path. The major parameter here is the updraft velocity, which is proportional to the eddy diffusion coefficient. The mean diameter of the particle estimated by this formula ranges in 0.1 -- 10 $\mu$m, depending on the ambient pressure and the updraft velocity, and this range is consistent with the values often adopted in the literature. More sophisticated, and potentially more realistic models of cloud particle sizes have been developed in Earth and planetary science context, and these models start to be used in exoplanet studies \citep[e.g.,][]{lavvas2017aerosol,gao2018sedimentation,kawashima2018theoretical}. While there are more detailed cloud microphysics models available, a simplified treatment would sufficiently answer our focused objectives as to whether and where condensation may occur and is thus used in this study.

\begin{figure}[ht]
\begin{center}
 \includegraphics[width=0.5\textwidth]{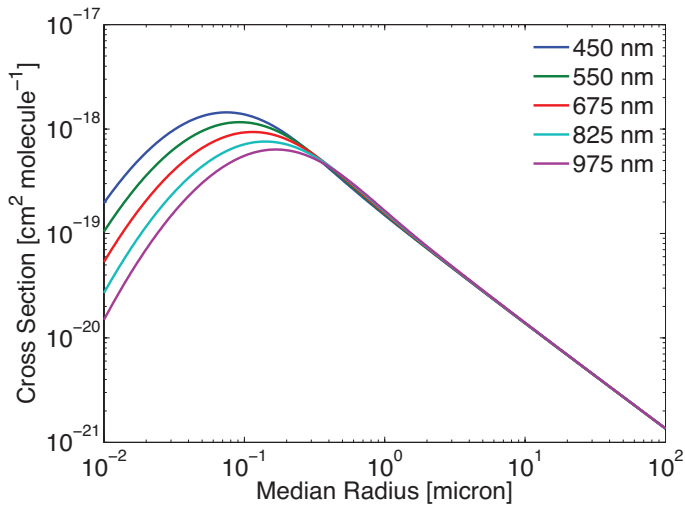}
 \caption{
The cross section of water molecule in the condensed phase as a function of the particle size. The horizontal axis is the median size or the zero-order moment of the lognormal size distribution. We assume the size dispersion parameter to be 2 in this calculation. Here we show the result for water ice, but the result for water liquid or ammonia ice is not substantially different. This figure shows that the radiative effect of condensed species is sensitive to the particle size of the cloud droplets and is maximized when the particle size is comparable to the wavelength. For particles larger than 1 $\mu$m, their cross sections in the visible-wavelength range are constant.
 }
 \label{fig:cross}
  \end{center}
\end{figure}

The cross section of cloud particles is calculated by the Mie theory. We assume a lognormal distribution with a dispersion factor of 2 \citep{Hu2013}. The calculation is performed for a mean particle radius ranging from 0.01 to 100 $\mu$m, which covers the range of interest. The refractive index of water ice is from \cite{Warren2008}, that of liquid water is from \cite{Hale1973}, and that of ammonia ice is from \cite{Martonchik1984}. We verify that at 500 nm the water droplet is almost fully reflective, with the single scattering albedo that only deviates from unity by less than $10^{-5}$. Figure \ref{fig:cross} illustrates that the radiative effect of cloud droplets is quite sensitive to the particle size. 

\subsection{Reflected Light Spectra} \label{sec:spectra}

The atmospheric pressure-temperature profile, mixing ratio profiles of molecules, and the clouds' radiative properties (i.e., opacity, single scattering albedo, and phase asymmetry factor) are used to compute the albedo spectrum of the planet at a specific phase angle of observation. The computation takes two steps. The first step is to compute the outgoing radiance from each illuminated patch of the atmosphere, and the second step is to integrate the results from the first step to obtain the albedo spectrum.

For the first step, we use the molecular line lists and collision-induced absorption opacities in the HITRAN 2012 database \citep{rothman2013hitran2012}. Specifically we include the molecular opacities of \ce{CH4}, \ce{NH3}, and \ce{H2O}, as well as the \ce{H2}-\ce{H2}, \ce{H2}-\ce{H}, and \ce{H2}-\ce{He} collision-induced opacities. The cross sections of \ce{CH4} at the visible wavelengths are taken from \citet{Karkoschka1998}. We calculate the molecular opacities with a line-by-line method, using the Voigt line profile, which incorporates both Lorentzian pressure broadening and Doppler broadening. We use the two-stream method for multiple scattering, with the $\delta$-Eddington approximation for starlight radiation \citep{Toon1989}. We then use the two-stream solution as the source function \citep{Toon1989} and solve the radiative transfer equation once again to obtain the outgoing radiance for a specified incident and emission angle pair. This last step also incorporates the component of single scattering, for which we use the exact phase function for Rayleigh scattering and the two-term Henyey-Greenstein phase function for cloud particle scattering. This treatment of the single-scattering component is the same as \cite{Cahoy2010}. 

We then integrate over the planetary disk by sampling the longitude and latitude with an 8-order Gaussian integration. For each specified phase angle of observation, the integration is performed over all latitudes and those longitudes that are illuminated by the starlight. Therefore, the sampling of the longitude (8 points) changes with the phase angle, and the sampling of the latitude (also 8 points) does not change. Each longitude-latitude pair corresponds to an incident and emission angle pair. In total, 64 plane-parallel calculations are performed and their results are then integrated to obtain the albedo spectrum. We define the ``albedo'' ($A$) as
\begin{equation}
\frac{F_p}{F_*} = A\bigg(\frac{R_p}{a}\bigg)^2, \label{albedo_definition}
\end{equation}
where $F_p$ and $F_*$ are the brightness of the star and the planet observed on Earth, and $R_p$ is the planetary radius, and $a$ is the planet's orbital distance from the star. $A$ thus contains both the factor of atmospheric scattering and absorption and also the phase angle of observation. When the phase angle is zero, such defined $A$ is the geometric albedo.

We test the radiative transfer calculations with idealized scenarios of semi-infinite homogeneous atmospheres whose geometric albedos have analytical solutions \citep{dlugach1974optical}. The scenarios include the Rayleigh and the Henyey-Greenstein phase functions with various values for the asymmetry factor and the single scattering albedo. We find excellent agreements between our model and \cite{dlugach1974optical}. An inter-model comparison has also been performed for several realistic planetary scenarios among several research groups (Mark Marley and Patrick Irwin, personal communications) as part of the WFIRST Coronagraph Data Challenge, and the comparison shows satisfactory agreement between the models.

An interesting phenomenon in the tested scenarios of semi-infinite homogeneous atmospheres is that the geometric albedos are very sensitive to trace amounts of absorption. A fully reflective (i.e., the single scattering albedo=1), isotropically scattering atmosphere would have a geometric albedo of 0.69; but the geometric albedo would be only 0.53 for a single scattering albedo of 0.99 \citep{dlugach1974optical}. For a Henyey-Greenstein phase function with an anisotropy parameter of 0.8, the geometric albedo further reduces to only 0.34 \citep[see also][Figure 4]{Lupu2016}. Therefore, we can expect that the ``continuum'' of the reflected light spectrum measures the interplay of the far wings of absorption features and the single scattering albedo and the degree of forward scattering of cloud particles.

\subsection{Jupiter as a Test Case} \label{sec:validation}

\begin{figure*}[ht]
\begin{center}
 \includegraphics[width=0.9\textwidth]{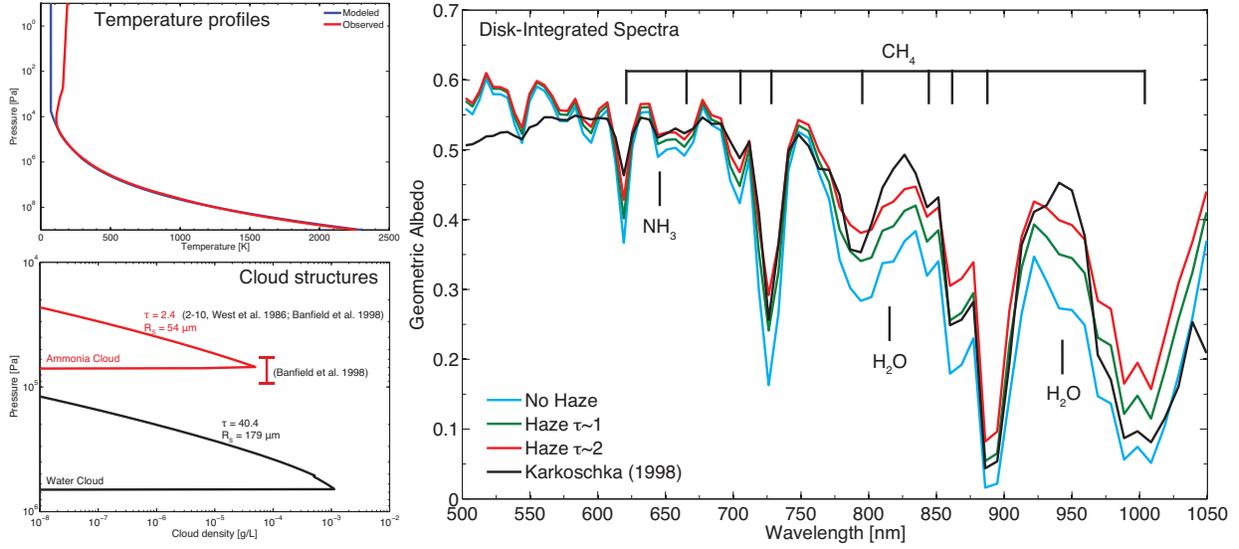}
 \caption{
Model results in comparison with Jupiter. The model is simulated for a Jupiter-mass, Jupiter-size exoplanet at 5.2 AU of a Sun-like star. The model assumes a 3x solar metallicity, an internal heat flux of 100 K, and an eddy diffusion coefficient of $10^4\sim10^6$ cm$^2$ s$^{-1}$. Varying the eddy diffusion coefficient by 2 orders of magnitude does not lead to appreciable changes to the results. The modeled temperature profile is consistent with the Galileo probe measurements and Cassini CIRS measurements \citep{Seiff1998, SimonMiller2006}, except in the stratosphere because heating due to \ce{CH4} and aerosol is not included in the model. For the cloud structure, the model accurately predicts the altitude and the vertical optical depth of the upper cloud layer made of ammonia ice, consistent with the telescopic, Galileo, and Cassini spectral retrieval \citep{West1986, Banfield1998, Matcheva2005}. The model can also produce an albedo spectrum that approximately matches the observed one \citep{Karkoschka1998}. The simple model slightly overestimates the weak bands of \ce{CH4} and the absorption of \ce{H2O}, which can be mitigated by adding a diffuse haze layer above the upper cloud, corresponding to Jupiter's upper-tropospheric haze \citep{Sato1979,West2004}.
 }
 \label{jupiter}
  \end{center}
\end{figure*}

We test our atmospheric cloud and reflected spectrum model for cold gas giants by simulating a Jupiter-sized planet at 5.2 AU of a Sun-like star. We adopt a 3x solar metallicity and an internal heat flux of 100 K as Jupiter \citep{Atreya2003,guillot2005interiors}. Figure \ref{jupiter} compares the modeled pressure-temperature profile, cloud structures, and albedo spectrum with the measured values.

This is an excellent agreement between the modeled and the measured temperature-pressure profiles for pressures greater than 0.1 bar, indicating that the model can correctly find where the adiabat would start. For pressures lower than 0.1 bar, the measured temperature profile shows a temperature inversion, but the modeled profile does not, and the modeled temperatures are substantially lower than the measured ones. This is because the stratosphere of Jupiter is additionally heated by absorption of \ce{CH4} and aerosols \citep{West1992, Moreno1997}, which is not included in the model. This lack of temperature inversion does not impede the model's ability to predict the cloud structures. For Jupiter, the model can predict the pressure of the ammonia ice cloud well consistent with the observed value (Figure \ref{jupiter}). 

The modeled albedo spectrum matches well with the observed disk-integrated spectrum in methane's strong bands but overestimates the weak bands of \ce{CH4} and the absorption of \ce{H2O} (Figure \ref{jupiter}). We think that the remaining discrepancy is mostly due to the upper tropospheric haze known to exist in the atmosphere of Jupiter \citep{Sato1979,Banfield1998}. This haze layer is not mainly composed of ammonia ice because ammonia ice features are not detected in the infrared spectra \citep{West2004}, and the leading candidate is hydrazine (\ce{N2H4}) produced from photolysis of ammonia. Since this photochemical haze is not produced in the model, we include it in an ad hoc way to explore its radiative effects. The haze is assumed to have a constant mixing ratio above the uppermost cloud deck for a vertical span of two scale heights, and is assumed to be white and has an extinction coefficient the same as ammonia ice. The mixing ratio of haze is adjusted to obtain the desired total optical depth of the haze layer. Figure \ref{jupiter} shows that the upper tropospheric haze can significantly affect the geometric albedo in weak methane bands. A thin, purely reflective haze layer having an optical depth of $1\sim2$ improves the consistency between the modeled and the observed albedo spectrum. 

In sum, the equilibrium cloud and reflected spectrum model can predict the pressures of the ammonia and water clouds, and the strengths of absorption bands in the reflected light spectra reasonably well. The model takes the irradiation flux, the internal heat flux, the atmospheric metallicity, and the surface gravity as the input parameters, and can compute the reflected light spectra at any phase angle of observation. The model captures the causal relationship between the input parameters and the pressure-temperature profiles, as well as the causal relationship between the pressure-temperature profiles and the condensation of water and ammonia in the atmosphere. The simple model is predictive and permits to explore a new scenario within seconds. Not included in the model is any ``puffy'' cloud structure caused by long-range vertical transport of condensate particles \citep{Ackerman2001}, any photochemical haze, or any additional absorbers in the clouds \citep{Wong2000, Wong2003}. With these caveats in mind, the current model is suitable for exploring the range of potential features in reflected light spectra of extrasolar giant planets and determining what we could learn from the spectra.

\section{Results}

\subsection{Cloud Type and Pressure Level} \label{sec:cloudtop}

We have simulated a grid of model scenarios that cover varied stellar irradiation flux, internal heat flux, atmospheric metallicity, and surface gravity. Because the model is not sensitive to the spectral shape of the host star, we always assume a Sun-like star and use the orbital distance expressed in the unit of AU as the measure of the stellar irradiation flux. We have explored 1.4 AU (ups And d), 1.7 AU (47 Uma b), 2.8 AU (Ups And e and 47 Uma c), and 3.8 AU (HD 160691 e), and they correspond to the irradiation fluxes received by the known wide-separation planets listed in parentheses. These planets, in particular, are often considered by direct imaging exoplanet missions \citep{Douglas2018spie,Seager2018SRM}. We have explored the surface gravity both at the low end ($10\sim25$ m s$^{-2}$, corresponding to Ups And e, 47 Uma c, and HD 160691 e) and at the high end ($60\sim100$ m s$^{-2}$, corresponding to ups And d and 47 Uma b). We use the 25 m s$^{-2}$ (corresponding to Jupiter) as the standard value, and discuss later the effects of higher surface gravity. Note that our ``high-end'' value of the surface gravity does not extend to planets more massive than 10 Jupiter's mass or brown dwarfs.

\begin{figure*}[ht]
\begin{center}
 \includegraphics[width=0.9\textwidth]{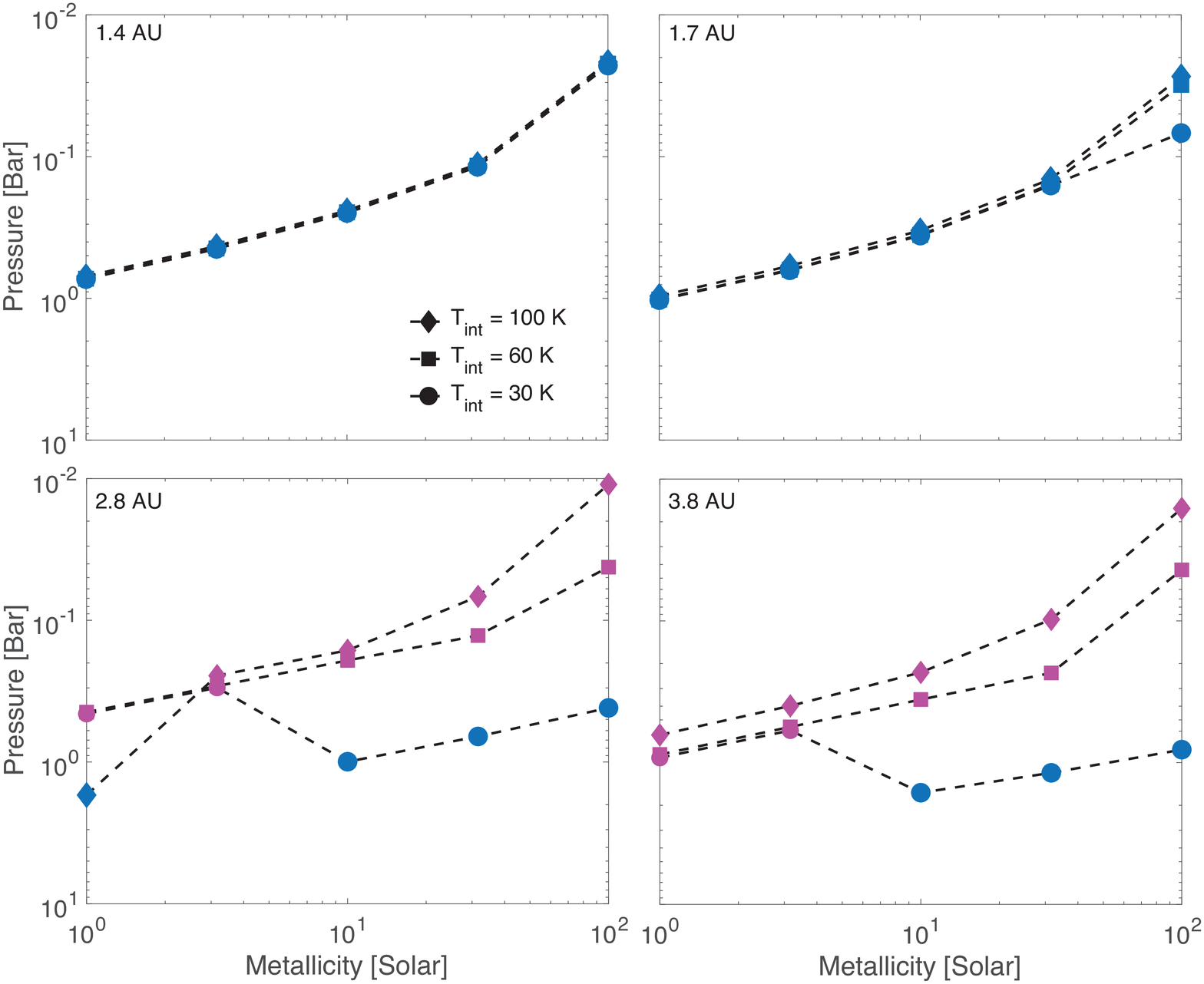}
 \caption{
Modeled cloud top pressure for a gaseous exoplanet at varied orbital distances from a Sun-like host star, as a function of the atmospheric metallicity. The atmosphere is dominated by \ce{H2} and \ce{He} and the surface gravity is 25 m s$^{-2}$. The cloud top pressure is defined to be the pressure where the cloud vertical optical depth equals to unity. The marker styles distinguish models using different internal heat fluxes. The markers are purple if the upper cloud deck is made of \ce{NH3} and the markers are blue if the upper cloud deck is made of \ce{H2O}.  The cloud top pressure is sensitive to the atmospheric metallicity and the internal heat flux.}
 \label{cloudtop}
  \end{center}
\end{figure*}

We explored a range of internal heat fluxes for each simulated planetary scenario, and the range corresponds to Jupiter ($T_{\rm int}=100$ K), Neptune ($T_{\rm int}=60$ K), and Uranus ($T_{\rm int}=30$ K). The internal heat flux of a given planet is not a free parameter, because it can be estimated by modeling the planet's thermal and structural evolution \citep[e.g.,][]{marley2007luminosity,fortney2007planetary,thorngren2018bayesian}. For example, the Jupiter-like internal heat flux we assumed may correspond to a Jupiter-mass planet of more than a billion years old, while a similarly old 10-Jupiter-mass planet would have an internal heat flux corresponding to $T_{\rm int}\sim300$ K \citep{marley2007luminosity}. Giant planets younger than 1 billion years old can have much higher internal heat fluxes \citep{marley2007luminosity}. Using the thermal evolution calculations, \cite{Lupu2016} estimated the internal heat fluxes for known widely separated giant planets, and found that very massive planets such as Ups And d could have high effective temperatures dominated by the internal heat flows. We additionally note that the internal heat flux predicted by thermal evolution models should probably be adopted as an upper limit, because there are mechanisms to temporarily inhibit vertical heat transport in giant planets \citep[e.g.,][]{leconte2017condensation}, and because empirically, the cause for Uranus's particularly low internal heat flux is still poorly understood \citep[e.g.,][]{fortney2010interior}. 

Figure \ref{cloudtop} shows the cloud top pressure, defined as the pressure at which the vertical optical depth of cloud particles equals unity, as a function of the atmospheric metallicity. Figure \ref{cloudtop} also indicates the type of the uppermost layer of clouds. The general trend is that when the planet is closer to the host star, the cloud is at a lower pressure; and when the atmosphere is more metal-rich, the cloud is at a lower pressure. These trends are consistent with the trends identified in previous works \citep[e.g.,][]{Sudarsky2003,Cahoy2010,Lupu2016,MacDonald2018}. For a giant planet at 1.4 and 1.7 AU from a Sun-like star, its uppermost cloud layer is made of \ce{H2O}, whereas planets in wider orbital separations may have an uppermost cloud layer made of \ce{NH3}. These trends and features are consistent with previous models of widely separated giant exoplanets \citep[e.g.,][]{Cahoy2010}.

\begin{figure*}[ht]
\begin{center}
 \includegraphics[width=0.9\textwidth]{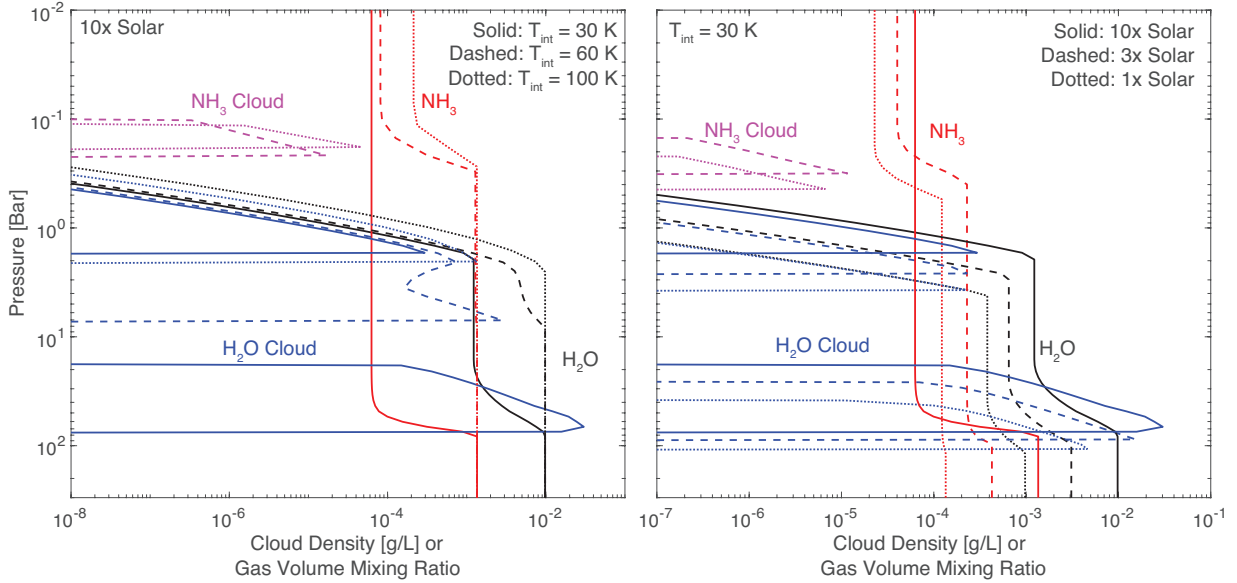}
 \caption{
Modeled atmospheric gas and cloud abundance profiles for a gaseous exoplanet at 2.8 AU from a Sun-like host star. The panels compare models for varied internal heat flux (left) and atmospheric metallicity (right). A thick, liquid water cloud forms when the internal heat flux is low and the atmospheric metallicity is moderately high, and this cloud absorbs and depletes \ce{NH3} in the atmosphere.
}
 \label{cloudexample}
  \end{center}
\end{figure*}

\begin{figure}[ht]
\begin{center}
 \includegraphics[width=0.45\textwidth]{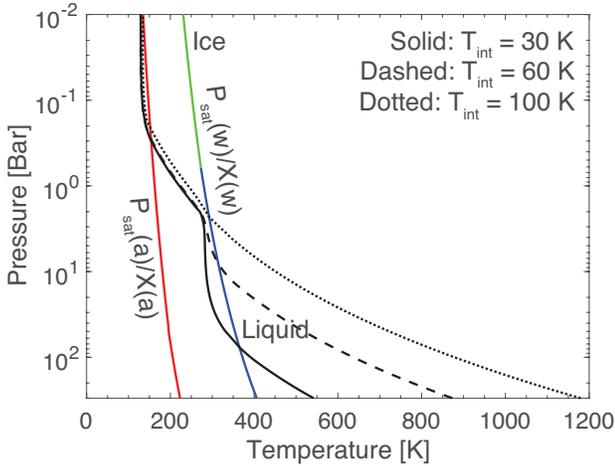}
 \caption{
Pressure-temperature profiles for a gaseous exoplanet at 2.8 AU from a Sun-like host star. The models have an atmospheric metallicity of $10\times$ Solar abundance and varied internal heat fluxes, corresponding to the left panel of Figure \ref{cloudexample}. Colored lines show the saturation vapor pressures of water and ammonia scaled by their pre-condensation mixing ratios, and green and blue are used to distinguish ice and liquid. The cloud condensation level of the $T_{\rm int}=30$ K case is deep in the atmosphere and that generates a massive liquid-water cloud.}
 \label{cloudtp}
  \end{center}
\end{figure}

A significant and new feature we find is that the cloud pressure is sensitive to the internal heat flux. Generally, when the planet has a greater internal heat flux, the cloud top is at a lower pressure. When the planet is as close as $1.4\sim1.7$ AU to the star, the internal heat flux has a minimal impact. When the planet is at $2.8\sim3.8$ AU from the star, however, the internal heat flux has a major impact. The uppermost cloud layer is \ce{NH3} for the moderate to high internal heat flux ($T_{\rm int}=60\sim100$ K), and that becomes \ce{H2O} for the small internal heat flux ($T_{\rm int}=30$ K) and a metallicity greater than approximately $10\times$ Solar  (Figure \ref{cloudtop}). The \ce{H2O} cloud as the uppermost cloud layer is deeper than the \ce{NH3} cloud as the uppermost cloud layer by a factor of $\sim5$ in pressure.

Why does the uppermost cloud layer become \ce{H2O} in some cases of the widely separated planets? A detail inspection of the model indicates that the formation of the \ce{NH3} cloud is prevented in these cases by the dissolution of \ce{NH3} into a deep water cloud in the form of liquid water droplets. Figures \ref{cloudexample} and \ref{cloudtp} show that a deep water cloud forms at 10 -- 100 bars when $T_{\rm int}=30$ K. This deep water cloud has liquid droplets more than $10^{-2}$ g/L, and the liquid droplets absorb \ce{NH3} in the atmosphere and deplete \ce{NH3} gas above the cloud. This ammonia depletion is preserved to the upper atmosphere by vertical mixing and eventually leads to the \ce{NH3} cloud not being able to form. The atmospheric metallicity impacts the ability of the deep liquid water cloud to absorb \ce{NH3}. The cloud density is higher for a greater mixing ratio of water in the deep atmosphere. Because \ce{NH3} is partitioned between the dissolved phase and the gas phase, a higher cloud density means that a greater fraction of \ce{NH3} is partitioned into the dissolved phase. The right panel of Figure \ref{cloudexample} shows that when the metallicity is $10\times$ Solar, the depletion of the \ce{NH3} into the thick water cloud at $\sim100$ bars is particularly severe, to the point that the \ce{NH3} cloud can no longer form in the upper atmosphere. This deep water cloud also forms for the planet at 1.7 AU having a $100\times$ Solar metallicity (Figure \ref{cloudtop}, the upper-right panel), except that in that case, the \ce{NH3} depletion is not visible via the cloud pressure as the \ce{NH3} cloud does not form even without the depletion.

One might ask why the $T_{\rm int}=60\sim100$ K cases in Figure \ref{cloudtp} do not have significant ammonia depletion given that their pressure-temperature profiles also intersect with the liquid part of the water saturation vapor pressure profile. This is because the cloud density must reach a critical value before \ce{NH3} is significantly partitioned in the liquid phase. The reduction factor for \ce{NH3} in the gas phase ($f$), defined as the ratio between the mixing ratio of \ce{NH3} in equilibrium with the cloud droplets and that in total, is
\begin{equation}
f = \frac{1}{1+\frac{RTH(1+K/[\ce{OH^-}])}{101325}\rho_c}, \label{partition}
\end{equation}
where $H$ is Henry's law constant in M atm$^{-1}$, $K$ is the equilibrium constant for ammonia dissolution in water in M, and [\ce{OH^-}] is the molality of \ce{OH-} in water, and the cloud density $\rho_c$ is in g/L. Using relevant values at the temperature of 300 K, Equation (\ref{partition}) becomes $f=1/(1+260\rho_c)$, and therefore the cloud density must be greater than approximately $10^{-2}$ g/L to cause a significant reduction in the \ce{NH3} mixing ratio in the gas phase. Even though the $T_{\rm int}=60\sim100$ K cases have liquid droplets at the base of the water clouds, they cannot significantly sequester \ce{NH3} because the cloud densities are too small. This sensitivity on the water cloud density also explains the dependency on the atmospheric metallicity seen in the right panel of Figure \ref{cloudexample}. The partitioning also depends on temperature. As the temperature increases, $H$ decreases and $K$ increases for \ce{NH3} . The change in $K$ dominates, and thus more \ce{NH3} is partitioned in the liquid phase when the temperature increases. In all, the effect of \ce{NH3} dissolution is more significant when more water condenses to the liquid form in the deeper and hotter part of the atmosphere.

The deep liquid-water cloud and \ce{NH3} depletion is more prevalent for a greater surface gravity. When using a surface gravity of $60\sim100$ m s$^{-2}$ for the planet at $2.8\sim3.8$ AU, we find that all cases with $T_{\rm int}=30$ K feature the deep cloud and miss the \ce{NH3} top cloud. Even at the Solar metallicity, the deep cloud can deplete \ce{NH3}. The overall cloud top pressure thus also has a sensitivity on the surface gravity.

In addition to the major sensitivity to the internal heat flux caused by \ce{NH3} dissolution described above, another interesting feature to note is that an optically thick ammonia cloud may not form if the atmospheric metallicity is not high enough. This is seen for a planet at 2.8 AU having a Solar metallicity (Figure \ref{cloudtop}, the lower-left panel). The case for a high internal heat flux ($T_{\rm int}=100$ K) shows the uppermost cloud layer to be \ce{H2O} and quite deep. A very thin \ce{NH3} cloud is formed in this case, but its optical depth is too small to have a meaningful impact on the spectrum.

Collectively, these results demonstrate that the cloud top pressure does not only depend on the molecular composition or the irradiation level but also significantly depends on the internal heat flux and to a lesser extent the surface gravity. Previous works additionally show that the cloud top pressure depends on the degree of atmospheric mixing \citep[e.g.,][]{MacDonald2018}. The potential to form a liquid water cloud deep in the atmosphere and the interaction of this cloud with \ce{NH3} significantly amplified the dependency on the internal heat flux and the surface gravity. 

\subsection{Features in the Reflected Light Spectrum} \label{sec:albedo}

\begin{figure}[ht]
\begin{center}
 \includegraphics[width=0.45\textwidth]{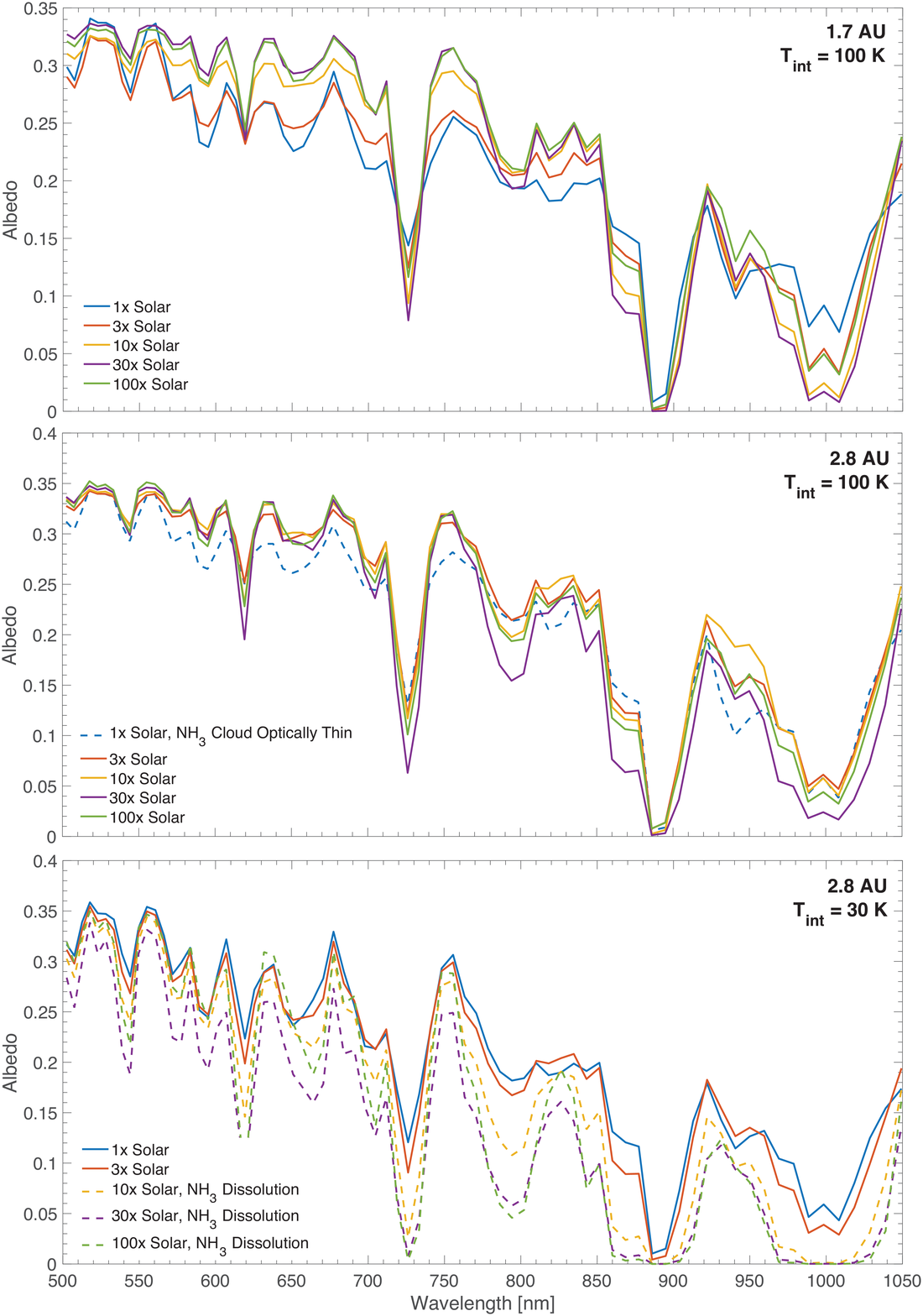}
 \caption{
Albedo spectra at the phase of $\pi/3$ for the modeled planetary scenarios. The albedo is defined in Equation (\ref{albedo_definition}), and other aspects of the models are presented in Figures \ref{cloudtop} and \ref{cloudexample}. The \ce{NH3} dissolution has a large impact on the albedo spectrum.
}
 \label{albedo}
  \end{center}
\end{figure}

Both the cloud top pressure and the mixing ratio of \ce{CH4}, and other species to a lesser extent determine the features in the reflected light spectrum. Figure \ref{albedo} shows the albedo spectra at the phase of $\pi/3$ for several representative cases. The cloud top generally moves up for a higher metallicity. Meanwhile, the atmosphere has more \ce{CH4}. The top panel in Figure \ref{albedo} (i.e., the 1.7-AU planet) shows both of these two factors at work: the strong bands of \ce{CH4} become deeper for a higher metallicity, because they depend more on the \ce{CH4} abundance; while the week bands and the continuum become shallower, because they depend more on the cloud pressure. This trend is reversed for the high metallicity of $100\times$ Solar, as in that case, the cloud top pressure decreases a lot faster than the metallicity increases (Figure \ref{cloudtop}, upper-right panel). The 1.7-AU planet scenarios that have the water cloud as the uppermost cloud layer can be compared with the models presented in \cite{MacDonald2018}. Similar to what's presented here, \cite{MacDonald2018} found that as the metallicity increases, the methane features strengthen and the continuum brightens. The underlying driver for these trends may however be different, as in our case the change is partly driven by the cloud moving up in the atmosphere, while in \cite{MacDonald2018} the change was partly driven by the cloud becoming thicker.

The \ce{NH3} dissolution has a large impact on the reflected light spectrum. The bottom panel of Figure \ref{albedo} (the 2.8-AU planet) shows that the \ce{CH4} absorption features become wide and deep, and in many cases saturated, when the \ce{NH3} dissolution occurs. Without the \ce{NH3} top cloud, the uppermost cloud layer would be at $\sim1$ bar for a $10\times$ Solar metallicity (Figure \ref{cloudtop}, lower-left panel), and this deep cloud and moderately high abundance of \ce{CH4} cause the strong absorption in the albedo spectrum. Comparing the $3\times$ and $10\times$ Solar cases, the impact of the \ce{NH3} dissolution and lack of \ce{NH3} cloud is darkening of the whole spectrum and broadening of the \ce{CH4} absorption features. These changes are well within the potential sensitivity of future exoplanet direct imaging experiments.

In addition to the absorption features of \ce{CH4}, many models in Figure \ref{albedo} show the absorption feature \ce{H2O} at $\sim940$ nm. This feature is weaker compared to the nearby \ce{CH4} features because \ce{H2O} is severely depleted by condensation in the observable part of the atmosphere. This \ce{H2O} feature becomes invisible when the nearby \ce{CH4} features are too wide. The models presented here confirm the suggestion of the absorption features of \ce{H2O} for widely separated planets hotter than Jupiter \citep{MacDonald2018}. The strength of the \ce{H2O} feature at $\sim940$ nm is greater for the 1.7-AU planet than the 2.8-AU planet, and only has a weak dependency on the metallicity (see Figure \ref{albedo}). This is also consistent with \citet{MacDonald2018}, who argued that the prominence of the \ce{H2O} feature instead mainly depends on the sedimentation efficiency of cloud particles.

With the strong and weak bands of \ce{CH4} and the feature of \ce{H2O}, none of the models in Figure \ref{albedo} are degenerate with each other for the entire wavelength range. Nonetheless, spectral degeneracy would arise if the wavelength coverage is limited. For example, if we look at the strong \ce{CH4} band at 725 nm alone, the $10\times$ and $30\times$ Solar cases for the 1.7-AU planet appear to be quite similar (Figure \ref{albedo}, top panel). Also, in the wavelengths $>750$ nm, the $30\times$ and $100\times$ Solar cases for the 2.8-AU planet with $T_{\rm int}=30$ K are nearly identical (Figure \ref{albedo}, bottom panel). These two degenerate cases are distinguishable at other wavelengths, e.g., between 600 and 700 nm, via different strengths in the weak bands of \ce{CH4}.

In sum, the \ce{CH4} absorption features in the reflected light spectrum are mainly sensitive to the cloud top pressure and the mixing ratio of \ce{CH4}, which in terms are controlled by the atmospheric metallicity and the internal heat flux of the planet. \ce{NH3} dissolution into a deep liquid-water cloud creates a major impact on the cloud top pressure and the reflected light spectrum. Both the strong and the weak bands of \ce{CH4} are needed to measure the cloud top pressure and the mixing ratio. Besides \ce{CH4}, \ce{H2O} has a secondary impact to the spectrum.
 
\subsection{Retrieval of the Cloud Pressure and the Mixing Ratio of Methane}

Up to this point, we have assumed that the abundance of \ce{CH4} is fully consistent with the atmospheric metallicity and a solar C/O ratio. This assumption should probably be considered as a ``weak prior'' when retrieving the atmospheric abundances from the reflected light spectra. In this section, we loose this constraint to see whether the cloud top pressure and the methane abundance can be simultaneously measured from a single reflected light spectrum.

\begin{figure*}[ht]
\begin{center}
 \includegraphics[width=0.9\textwidth]{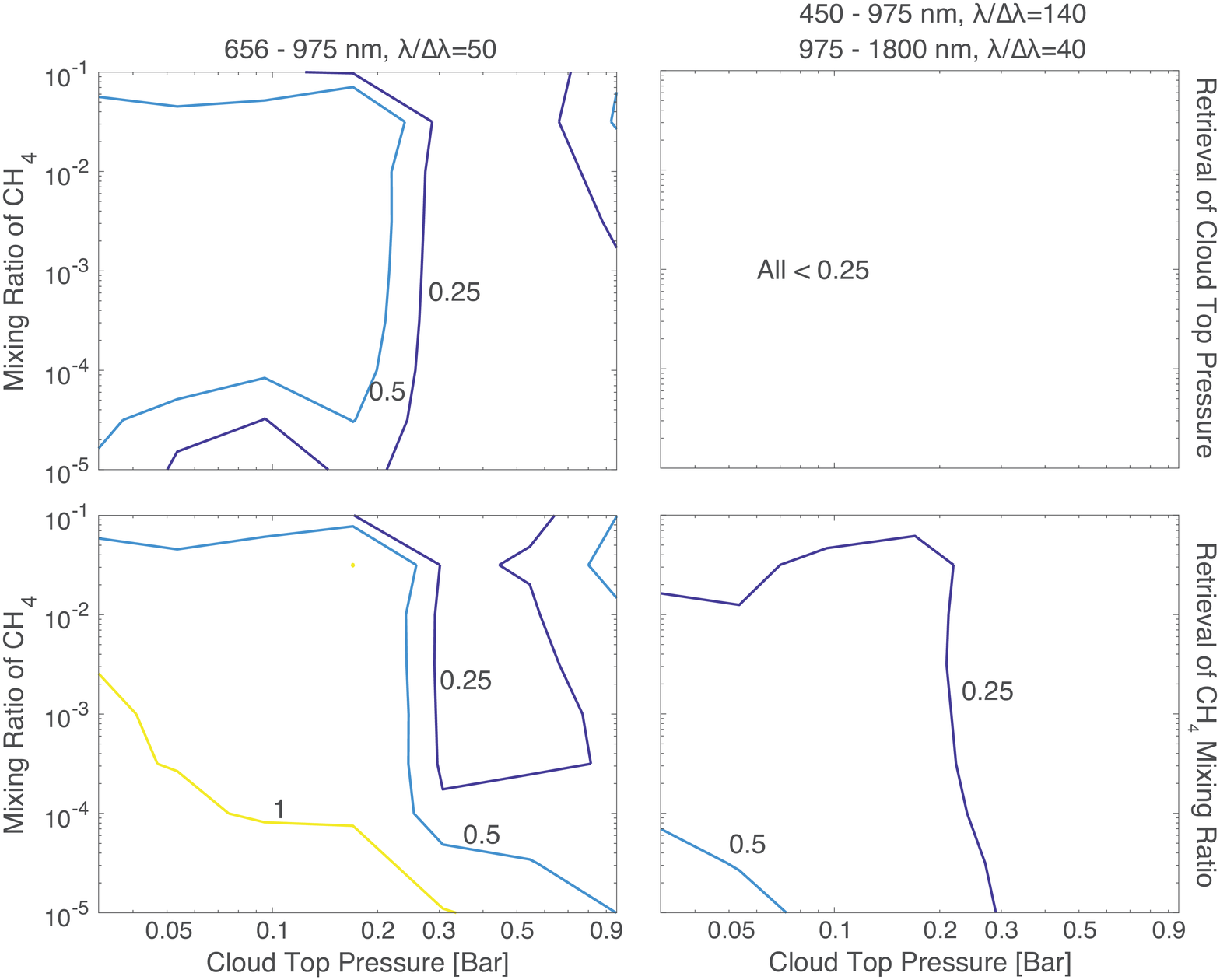}
 \caption{
Expected uncertainties in the retrieved cloud top pressure and mixing ratio of \ce{CH4} in the dec unit. The posterior is derived linearly (Eq. \ref{covariance}) from the Jacobian evaluated at each grid point and the measurement error assumed to be a constant corresponding to S/N=20 at the wavelength of 600 nm. The expected uncertainty is $<0.25$ dec for both the cloud top pressure and the mixing ratio of \ce{CH4} when the cloud is deeper than $\sim0.3$ bar and \ce{CH4} is more abundant than $\sim10^{-4}$.  
}
 \label{retrieval}
  \end{center}
\end{figure*}

To study the retrieval of these two key parameters, we set up a grid of models that explore the cloud top pressure and the mixing ratio of \ce{CH4}. Using the 1.7-AU planet with $T_{\rm int}=100$ K as the test case, we obtain a grid of model with the cloud top pressure from 0.96 to 0.03 bar, equally spacing in the logarithmic scale by 0.25 dec, for a metallicity grid of 1.0, 3.9, 11.7, 25.7, 45.7, 69.2, 93.3$\times$ solar abundance. For each scenario, we calculate a set of reflected light spectra at the phase of $\pi/3$ that explore the mixing ratio of \ce{CH4} ranging from $10^{-5}$ to $10^{-1}$, equally spaced in the logarithmic scale by 0.5 dec.

Rather than performing a full retrieval, we apply the information content approach in atmospheric remote sensing to obtain an understanding of the retrievability of the parameters. For each point in the grid, we calculate the Jacobian matrix (${\bf K}$), and use it to calculate the covariant matrix of the posterior (${\bf S}$)
\begin{equation}
{\bf S = (K^TS_{\epsilon}^{-1}K + S_a^{-1})^{-1}}, \label{covariance}
\end{equation}
where ${\bf S_{\epsilon}}$ is the measurement error covariance, and ${\bf S_a}$ is the covariance of the prior. This formula is well established in atmospheric remote sensing \citep{rodgers2000inverse}, and has been used to study transmission spectroscopy of exoplanets \citep{batalha2017information}. 

The covariant matrix of the posterior ${\bf S}$ is a $2\times2$ matrix in our problem, and it defines the probability density function of the posterior. The constant probability contour is an ellipse in the phase space, whose semi-major axis is the larger of the eigenvalues of ${\bf S}$ and the principal axis is the corresponding eigenvector. Let $\lambda$ be the eigenvalue and $v=(v_1,v_2)$ be the eigenvector, the uncertainty in the retrieved parameter is simply $C \lambda v_1$ and $C \lambda v_1$, where $C$ is a constant that we adopt to be 3 for a conservative estimate. We use 10-base logarithmic values of the cloud top pressure and the \ce{CH4} mixing ratio as the state vector, and as such the uncertainty in the retrieved parameter is dimensionless and in the unit of dec. In this work we assume that the prior is weak, i.e., ${\bf K^TS_{\epsilon}^{-1}K \gg S_a^{-1}}$. We also assume a constant noise for each spectral element, corresponding to S/N=20 at the wavelength of 600 nm. This assumption on noise is simple and a more realistic treatment can be found in \citet{robinson2016characterizing}. We consider two observational scenarios. The first one, corresponding to the Starshade Rendezvous Probe \citep{Seager2018SRM}, is a spectrum in 656 -- 975 nm at a resolution of 50. The second one, corresponding to HabEx \citep{mennesson2016habitable}, is a spectrum in 450 -- 1800 nm having a resolution of 140 in 450 -- 975 nm and 40 in 975 -- 1800 nm. Note that while the spectral coverage and resolution correspond to these concepts, we assumed a measurement error constant as a function of wavelength and defined the S/N at 600 nm in this analysis. The current analysis does not include any additional instrument effects, such as the wavelength-dependent efficiency of detectors. The uncertainties in the retrieved parameter for these observational scenarios are shown in Figure \ref{retrieval}.

We find that both the cloud top pressure and the mixing ratio of \ce{CH4} can be determined from a single reflected light spectrum to better than 0.25 dec when the cloud is deeper than $\sim0.3$ bar and \ce{CH4} is more abundant than $\sim10^{-4}$. For the shallower cloud, the uncertainty in the parameters is larger and will be approximately $0.5\sim1$ dec for a spectrum of S/N=20 in 656 -- 975 nm. When the wavelength coverage extends to 450 -- 1800 nm (i.e., the HabEx scenario), the uncertainties in the retrieved parameters significantly improve. For the entire parameter space explored (i.e., the cloud top pressure in 0.03 -- 1 bar, and the \ce{CH4} mixing ratio in $10^{-5}\sim10^{-1}$), the cloud top pressure can be measured with a posterior distribution narrower than 0.25 dec, and the \ce{CH4} mixing ratio with a distribution narrower than 0.5 dec.

\begin{figure}[ht]
\begin{center}
 \includegraphics[width=0.45\textwidth]{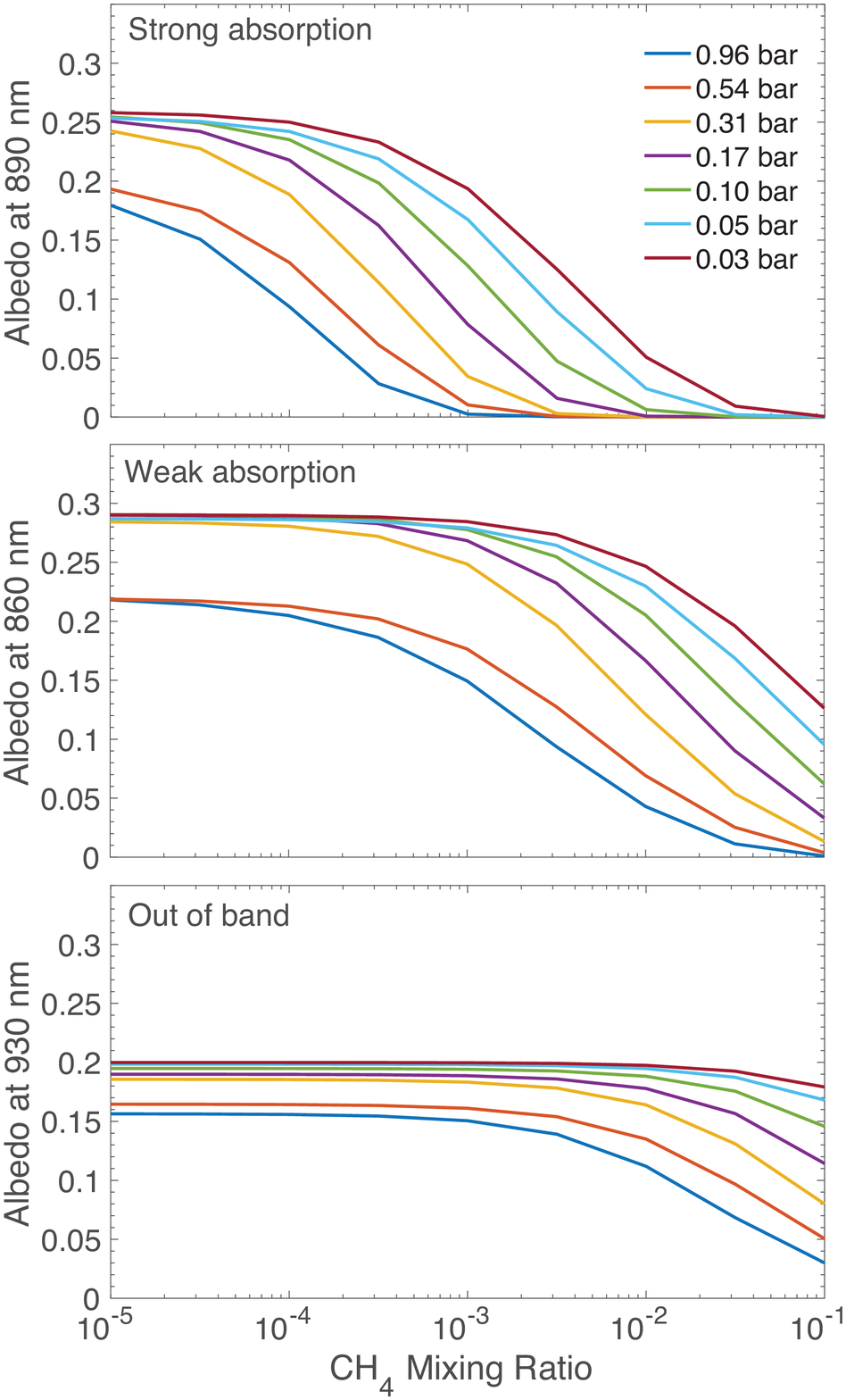}
 \caption{
The growth curves of \ce{CH4} absorption features for varied cloud top pressures. The three panels correspond to the center of a strong \ce{CH4} absorption band, an adjacent weaker absorption band, and a wavelength out of any main absorption band. The growth curves are vastly different between the strong and the weak bands.
}
 \label{absorption}
  \end{center}
\end{figure}

We thus suggest that the cloud top pressure as deep as $0.3\sim1$ bar would be a favorable parameter range for measuring the cloud top pressure and the mixing ratio of \ce{CH4} from a single reflected light spectrum in wavelengths $<1\mu$m. The reason for this ``sweet spot'' is the relative strength between the strong bands and the weak bands of methane, as well as those spectral elements that do not contain any major absorption of \ce{CH4} (Figure \ref{absorption}). The strong and weak bands depend on the methane column differently, and the out-of-band albedo is mostly controlled by the cloud top pressure. As a well-known approach \citep[e.g.,][]{Marley2014report,Burrows2014report,Hu2014report}, these pieces of information can be used in the retrieval to measure the mixing ratio of \ce{CH4} and the cloud top pressure simultaneously. Notably, the growth curves of the weak band are better separated for the cloud top pressure in $0.3\sim1$ bar than the cases of lower cloud top pressures (Figure \ref{absorption}). In other words, when the uppermost cloud deck is deeper than $\sim0.3$ bar, the weak bands of \ce{CH4} are well developed depending on the mixing ratio of \ce{CH4}, and provide diagnosing power to break potential degeneracies between the mixing ratio and the cloud top pressure. When the cloud is higher, the weak bands are not sufficiently deep and provide less information. While not shown, we have performed simulations and found that when the cloud top pressure is deeper than $\sim1.5$ bars, the continuum is very low and many methane absorption features do not prominently show up in the spectrum, diminishing their diagnosing power. Therefore, the scenario with a cloud top pressure ranging in $0.3\sim1$ bar would have both strong and weak bands of \ce{CH4} well developed based on a high continuum, and thus is the favorable parameter range for determining the atmospheric abundances.

To summarize, when the uppermost cloud deck is deeper than $\sim0.3$ bar, the weak bands and strong bands of methane allow measurement of the \ce{CH4} mixing ratio and the cloud top pressure with one observation of the reflected light spectrum at wavelengths $<1\mu$m with S/N of 20. When the uppermost cloud deck is shallower than $\sim0.3$ bar, the methane mixing ratio and the cloud top pressure are somewhat correlated, leading to a measurement uncertainty for both parameters $>0.5$ dec. The measurements would be improved by extending the wavelength coverage to $1.8\ \mu$m. The results presented here are based on the linear formulation of the retrieval problem and assume a Gaussian posterior. The actual retrieval would require sampling of the posterior and may deviate from these assumptions significantly. The full retrieval based on ExoREL is described in a companion paper (Damiano \& Hu 2019, submitted).

\section{Discussion}

\subsection{Implications for Future Direct Imaging Experiments}

The analysis in this paper shows that reflected light spectra of cold giant exoplanets can be used to determine the abundance of methane in their atmospheres, as well as the pressure of the uppermost cloud layer. In general, S/N of 20 is required to yield meaningful constraints (Figure \ref{retrieval}); we have tested smaller S/N, such as 10, and found that the constraints would deteriorate substantially. This is consistent with the previous analysis using retrieval \citep{Lupu2016}, and it is interesting to note that the required S/N for meaningfully characterize Earth-like planets in the reflected light is also $\sim20$ \citep{Feng2018}. The wavelength coverage and the spectral resolution also matter. If limited to the wavelengths $<1\mu$m and a spectral resolution of 50, such as the Starshade Rendezvous Probe in tandem with WFIRST, the planets that are best for atmospheric characterization will be those that have the uppermost cloud layer deeper than $\sim0.3$ bar.

This favorable parameter space for atmospheric characterization may inform the selection of targets for time-heavy integration to obtain the reflected light spectra. Comparing the 0.3 bar pressure criterion with Figure \ref{cloudtop}, a 1.7-AU planet would be a better target than a 1.4-AU planet in this regard, and a 3.8-AU planet would be a better target than a 2.8-AU planet. In other words, within the group of the ``water-cloud'' planets and the ``ammonia-cloud'' planets, the colder one would have a deeper cloud for the same atmospheric metallicity, and thus better developed strong and weak bands of \ce{CH4}, leading to better constraints on the posterior for the same S/N. The AU numbers describe the irradiation flux the planet receives and does not necessarily imply the orbital distance from the parent star.

Once the cloud pressure and the mixing ratio of \ce{CH4} are derived from the reflected light spectrum, one can use ExoREL to calculate the pressure-temperature profile and determine the type of cloud particles. Even though the atmospheric retrieval itself may not yield information regarding the composition of the cloud, we expect little ambiguity between \ce{NH3} clouds and \ce{H2O} clouds once the cloud top pressure is measured, because the pressures of these two types of clouds generally do not overlap for a given planet (Figure \ref{cloudtop}).

\subsection{Cloud Pressure as an Indicator of Thermal Evolution}

Section \ref{sec:cloudtop} shows that the internal heat flux has a significant bearing on the overall cloud top pressure of the atmosphere. This dependency is particularly strong between a low, Uranus-like heat flux and the higher fluxes, in that the low heat flux would result in a liquid-water cloud that dissolves and depletes \ce{NH3} deep in the atmosphere. Therefore, determining the cloud top pressure appears to provide a highly useful way to measure the internal heat flux of a giant exoplanet. Particularly for the 2.8-AU planet, the scenarios with \ce{NH3} dissolution have cloud top pressures $>0.4$ bar, while the scenarios without generally have higher clouds. From Figure \ref{absorption}, the growth curves of \ce{CH4} absorption are quite different between these two scenarios, indicating that \ce{NH3} dissolution caused by a low internal heat flux is well detectable by the reflected light spectrum. The internal heat flux is a key prediction of evolutionary models of giant planets \citep{Baraffe2003,Baraffe2008,Fortney2006}, but its direct measurement would, by definition, require a wide wavelength coverage well into in the thermal infrared. As such a wavelength coverage may not be feasible for many planets widely separated from their host stars, reflected light spectroscopy in direct imaging may offer an alternative way to study the evolution of giant exoplanets. Using the cloud pressure as an indicator of the thermal evolution would be particularly useful to distinguish a small internal heat flux, because that's when the dissolution of \ce{NH3} would occur, but less so to distinguish an internal heat flux larger than $T_{\rm int}=100$ K as predicted for planets younger or more massive than 10 Jupiter's mass.

\subsection{Other Impact of the Dissolution of \ce{NH3}}

The formation of a \ce{NH4SH} cloud via the reaction between \ce{NH3} and \ce{H2S} and subsequent condensation has been predicted to occur in the atmosphere of Jupiter \citep[e.g.,][]{Lewis1969,Weidenschilling1973} as well as in low-temperature planetary atmospheres \citep[e.g.,][]{lodders2002atmospheric}. Although our model does not include the \ce{NH4SH} cloud, here we discuss its impact. The formation of the \ce{NH4SH} cloud occurs at the pressure level of $0.1\sim10$ bars \citep{lodders2002atmospheric}, while the dissolution of \ce{NH3} in liquid-water clouds would occur at the pressure level of $30\sim100$ bars (Figure \ref{cloudexample}). Another fact to consider is that the cosmic abundance of N is greater than that of S by approximately a factor of 5. Therefore, if the atmosphere we model has a cosmic N/S ratio, the scenarios without the dissolution of \ce{NH3} could have  \ce{NH4SH} condensation, and the resulting \ce{NH4SH} cloud should sequester almost of the entirety of \ce{H2S}, just like in the atmosphere of Jupiter. This condensation only causes a mild reduction in the abundance of \ce{NH3} and does not prevent the formation of the \ce{NH3} cloud in most cases. Atmospheres warmer than the Jupiter's may not have the condition for \ce{NH4SH} condensation and thus may have \ce{H2S} and its photochemical haze in the upper atmosphere \citep{gao2017sulfur}. Our scenarios with the dissolution of \ce{NH3} would greatly deplete \ce{NH3} in the atmosphere, and it would prevent the formation of the \ce{NH4SH} cloud. In addition to the removal of the \ce{NH3} cloud as modeled in this paper, we expect the effects of the \ce{NH3} dissolution to also include the removal of the \ce{NH4SH} cloud. This would be yet another way to liberate \ce{H2S} to the upper atmosphere and cause photochemical hazes. 

\subsection{Caveats}

The determination of the cloud top pressure may be complicated by haze or partial cloud coverage. First, a potential photochemical haze would bias the inferred cloud top pressure if not considered. The effect of the photochemical haze is reducing the strengths of the weak bands of methane (Figure \ref{jupiter}), which makes the cloud top pressure appear to be higher. There is no clear way to remove the effect of the photochemical haze, because the haze may lack any specific spectral features, like the upper tropospheric haze of Jupiter. The investigation of the effect of a potential haze layer would likely rely on atmospheric photochemistry models \citep{Hu2012,Hu2013,lavvas2017aerosol,gao2018sedimentation,kawashima2018theoretical}. Second, the inference based on one-dimensional models would indicate a cloud top pressure on average. If the exoplanet has a banded cloud structure like Jupiter, the cloud top pressure derived would likely indicate a weighted average of that of the ``belts'' and that of the ``zones''. Because the zones would be brighter and contribute more reflected light than the belts, the one-dimensional model would likely find a value close to the cloud top pressure of the zones corresponding to the updraft portions of convective cells \citep{Ingersoll2004}. 

\section{Conclusion}

We present a new, one-dimensional, equilibrium cloud and reflected light spectrum model for widely separated gaseous exoplanets having \ce{H2}/He dominated atmospheres. The model, called ExoREL, takes the irradiation flux, the internal heat flux, the atmospheric metallicity, and the surface gravity as the input parameters, and can compute the reflected light spectra at any phase angle of observation. The model includes condensation of \ce{H2O} and \ce{NH3} to form respective cloud layers. The model also includes the dissolution of \ce{NH3} into water clouds when the cloud particles are liquid droplets, an effect modeled for exoplanets for the first time. ExoREL captures the causal relationship between the input parameters and the pressure-temperature profiles, as well as the causal relationship between the pressure-temperature profiles and the condensation of water and ammonia in the atmosphere. The simple model is thus predictive and permits to explore a wide range of parameters. The model is sufficiently fast so that it provides the basis for a Bayesian retrieval method for the reflected light spectrum (Damiano \& Hu 2019, submitted).

We have used the model to explore the range of potential features in reflected light spectra of giant exoplanets and to determine what we could learn from the spectra. By simulating model scenarios for varied irradiation flux, internal heat flux, and atmospheric metallicity, we identify two new findings. (1) We find that a low, Uranus-like internal heat flux would result in a layer of liquid-water cloud deep in the atmosphere. When the density of this liquid-water cloud is higher than $\sim10^{-2}$ g/L, this cloud would significantly deplete \ce{NH3} by dissolution and prevent the formation of the \ce{NH3} cloud that would otherwise be the uppermost cloud layer. The net result is a strong sensitivity of the type of the uppermost cloud layer (\ce{NH3} vs. \ce{H2O}), as well as the cloud top pressure, on the internal heat flux. (2) We also find that the atmospheres with a cloud top pressure ranging in $0.3\sim1$ bar would have both strong and weak bands of \ce{CH4} well developed in their reflected light spectra. These bands together provide the diagnosing power to determine the atmospheric abundance of \ce{CH4} and the cloud top pressure uniquely from a single observation of the reflected light spectrum at wavelengths $<1\ \mu$m with S/N=20. The atmospheres with a higher cloud top would likely require a wider wavelength coverage to characterize. 

Collectively, the findings presented in this paper reaffirm the richness of information that can be obtained from the reflected light spectra of widely separated giant exoplanets. With the cloud top pressure measured by the reflected light spectra, the planet's internal heat flux can also be inferred, and a planet with ``normal'', Jupiter-like internal heat flux can be distinguished from a planet with Uranus-like heat flux. The reflected light spectrum is thus a way to characterize not only the atmospheric abundances but also the thermal evolution of giant exoplanets. 

\acknowledgments
We thank Wesley Traub, Adam Burrows, Mark Marley, Jonathan Fortney, Yuk Yung, Sara Seager, Bruce Macintosh, Margaret Turnbull, and Mario Damiano for motivation, discussion, and support that enabled this work. This work was supported in part by the NASA WFIRST Preparatory Science grant \#NNN13D460T, NASA WFIRST Science Investigation Teams grant \#NNN16D016T, and NASA Exoplanets Research Program \#80NM0018F0612. The research was carried out at the Jet Propulsion Laboratory, California Institute of Technology, under a contract with the National Aeronautics and Space Administration.

\appendix

\section{Derivation of the Formula for Mean Particle Diameter} \label{app:particlesize}

Here we present the derivation of the formula for the mean particle diameter (Equation \ref{eq:particlesize}), with a simplified approach to estimate the particle size of aerosols (cloud and haze particles) in exoplanet atmospheres. Aerosols may be formed via homogeneous or heterogenous nucleation when the corresponding gas phase is supersaturated. Once formed, the aerosol particles may grow by condensation, or shrink by evaporation. The rate of condensation and evaporation is proportional to the difference between the actual gas phase concentration and the saturation concentration \cite[e.g.,][]{Seinfeld2006}. The aerosol particles may also collide with each other and merge to form larger particles. Finally, the aerosol particles settle downwards due to gravity. We formulate the growth of an aerosol particle population taking into account all these processes with some important simplification. The purpose of the simplified approach is to elucidate the most important factors that control the size of aerosol particles.

We assume that the particles obey a lognormal size distribution fully characterized by the total number concentration of particles ($N$), the mean volume of particles ($V$), and the size dispersion of the population ($\sigma$). The size dispersion depends on interactions of all internal timescales of nucleation and coagulation, so it requires full microphysical simulations to be determined. However, under atmospheric conditions on Earth and other Solar System planets, the size dispersion parameter usually takes a value between one and two \cite[e.g.][]{Jonsson1996,Seinfeld2006,Knollenberg1980}. We will in the following assume $\sigma$ to be 2. Note that mass added to the condensed phase by nucleation and coagulation is negligible; in mathematical terms, this corresponds to $NV={\rm const.}$ for nucleation and coagulation.

The evolution equation for the particle population is characterized by
\begin{eqnarray}
&& \frac{dN}{dt} = J - KN^2 - \frac{v_d}{H}N,\nonumber\\
&& \frac{dV}{dt} = -\frac{V}{N}J + KVN + {\cal D}V^{1/3},\label{eq1}
\end{eqnarray}
in which $J$ is the nucleation rate, $K$ is the coagulation coefficient, $v_d$ is the effective gravitational settling velocity, $H$ is the atmospheric scale height, and ${\cal D}$ is the effective condensation coefficient. The equation (\ref{eq1}) describes the evolution of an aerosol population that condenses from supersaturation: the first two terms in both equations represent nucleation and coagulation; the third term in the $N$ equation represents the combination of eddy mixing and gravitational settling; and the third term in the $V$ equation represents the condensational growth of each particle, the main mechanism that transfers mass from the gas phase to the condensed phase. Note that I have used the $NV={\rm const.}$ to derive the terms corresponding to nucleation and coagulation.

The coefficient for each physical processes in equation (\ref{eq1}) can be related to atmospheric parameters or material parameters as follows. The nucleation rate $J$ is little constrained; however we will see in the following that the mean particle size does not depend on it. Coagulation is related to the Brownian motion of particles, such that
\begin{equation}
K = \frac{4kT}{3\mu},
\end{equation}
where $k$ is the Boltzmann constant and $\mu$ is gas dynamic viscosity. Gravitational settling of particles is compensated by updraft, so that
\begin{equation}
v_d = \max \bigg[ \frac{\rho_{\rm p}gC_c}{(162\pi^2)^{1/3}\mu}V^{2/3} \exp(-\ln^2\sigma) - u , \ 0 \bigg],
\end{equation}
where $\rho_{\rm p}$ is the density of the condensed phase, $g$ is the gravitational acceleration, $C_c$ is the slip correction factor that approaches to unity if the particle size becomes much larger than the mean free path of the atmosphere,  and $u$ is the mean updraft velocity that can be related to the eddy diffusion coefficient $K_E$ as
\begin{equation}
u = \frac{K_E}{H}.
\end{equation}
The effective condensation coefficient ${\cal D}$ is
\begin{equation}
{\cal D} = \frac{(48\pi^2)^{1/3}Df_am(n-n_s)}{\rho_{\rm p}}\exp(-\ln^2\sigma),
\end{equation}
where $D$ is the diffusion coefficient of the gas phase, $f_a$ is the accommodation coefficient that again approaches to unity if the particle size becomes much larger than the mean free path of the atmosphere, $m$ is the molecular mass, $(n-n_s)$ is the number concentration of the gas phase in excess of saturation. See \cite{Seinfeld2006} for parameterizations for $C_c$ and $f_a$.

The Equation (\ref{eq1}) is analytically solvable for the steady state, in which $dN/dt=0$ and $dV/dt=0$. The solution of the steady-state mean particle volume comes from re-arranging Equation (\ref{eq1}) into the derivatives of total mass in the condensed phase $M\equiv NV$, such that
\begin{eqnarray}
\frac{dM}{dt} & = & \rho_{\rm p}\bigg( V\frac{dN}{dt} + N\frac{dV}{dt} \bigg) \nonumber \\
& = & \rho_{\rm p} \bigg[ - \frac{v_d}{H} NV + {\cal D}V^{1/3}N \bigg],
\end{eqnarray}
and then $dM/dt=0$ can be further re-arranged to a quadratic equation with respect to $V^{2/3}$:
\begin{equation}
-\frac{\rho_{\rm p}gC_c}{(162\pi^2)^{1/3}\mu H}\exp(-\ln^2\sigma)V^{2/3} + \frac{u}{H} + \frac{(48\pi^2)^{1/3}Df_am(n-n_s)}{\rho_{\rm p}}\exp(-\ln^2\sigma)V^{-2/3} = 0. \label{quad}
\end{equation}
Equation (\ref{quad}) implies that the steady-state mean particle volume (and size) only depends on the balance between condensational growth, gravitational settling, and updraft. One needs to know the values of $C_c$ and $f_a$ in order to solve Equation (\ref{quad}), which in turn depend on the particle size. Starting from $C_c=1$ and $f_a=1$, the self-consistent solution of the mean particle size can be found with a few iterations. The medium diameter useful in describing the lognormal distribution is
\begin{equation}
D_p = \bigg(\frac{6V}{\pi}\bigg)^{1/3}\exp\bigg(-\frac{3}{2}\ln^2\sigma\bigg),
\end{equation}
and the quadratic mean diameter, which is proportional to the particle's radiative cross section, is
\begin{equation}
\overline{D_S} = \bigg(\frac{6V}{\pi}\bigg)^{1/3}\exp\bigg(-\frac{1}{2}\ln^2\sigma\bigg).
\end{equation}

In addition to usual atmosphere and material parameters, the mean particle size solvable from equation (\ref{quad}) only depends on three free parameters: the updraft velocity $u$, the supersaturation $s\equiv(n/n_s-1)$, and the size dispersion ($\sigma$). The updraft velocity can be computed from 3-dimensional (3D) general circulation simulation for irradiated atmospheres \cite[e.g.,][]{Parmentier2013}, and the supersaturation of a certain gas can be computed from the chemistry-diffusion simulation of thermochemistry and photochemistry \citep[e.g.,][]{Hu2014}.  Equation (\ref{quad}) thus provides a practical way to estimate condensate particle size complementary to comprehensive exoplanet atmosphere models.

The solution of Equation (\ref{quad}) has two distinct parameter regimes. When updraft is relatively low, the main mechanism to compensate gravitational settling is condensation, so the regime is ``condensation-dominated"; when updraft is relative high, the main mechanism to compensate gravitational settling is updraft, so the regime is ``updraft-dominated". The division of the condensation-dominated regime and the updraft-dominated regime can be described by the following dimensionless parameter
\begin{equation}
{\cal K} = \frac{8gDm(n-n_s)\exp(-2\ln^2\sigma)}{3\mu u}. \label{regime}
\end{equation}
Here we have dropped the correction term $C_c$ and $f_a$. The aerosol particle population is in the condensation-dominated regime when ${\cal K}\gg1$ and in the updraft-dominated regime when ${\cal K}\ll1$. In the condensation-dominated regime, the mean particle volume at the steady state is
\begin{equation}
V_{\rm min} = 29.6\bigg[\frac{Df_am(n-n_s)\mu H}{\rho_{\rm p}^2gC_c}\bigg]^{3/4}. \label{smallu}
\end{equation}
This equation shows that a larger supersaturation can maintain larger particles, whereas denser materials tend to form smaller particles, reflecting the balance between condensation and settling. In the condensation-dominated regime, the mean volume no longer depends on the size dispersion. In the updraft-dominated regime, the mean particle volume at the steady state is
\begin{equation}
V_{\rm asym} = 39.9\bigg[\frac{\mu u \exp(\ln^2\sigma)}{\rho_{\rm p}gC_c}\bigg]^{3/2}, \label{largeu}
\end{equation}
which is independent from the supersaturation and scales with the updraft velocity as $V\propto u^{3/2}$. The quadratic mean particle diameter in the updraft-dominated regime no longer depends on the size dispersion, viz.,
\begin{eqnarray}
\overline{D_S} & = &   4.24 \bigg[\frac{\mu u }{\rho_{\rm p}gC_c}\bigg]^{1/2}, \nonumber \\
 & = & 0.95 C_c^{-1/2} \bigg(\frac{\mu}{10^{-5}\ {\rm Pa}\ {\rm s} }\bigg)^{1/2} \bigg(\frac{u}{10^{-4}\ {\rm m}\ {\rm s}^{-1} }\bigg)^{1/2}\nonumber\\
&&\times\bigg(\frac{\rho_{\rm p}}{2000\ {\rm kg}\ {\rm m}^{-3} }\bigg)^{-1/2}
\bigg(\frac{g}{10\ {\rm m}\ {\rm s}^{-2} }\bigg)^{-1/2} \ {\rm \mu m}. \label{scale}
\end{eqnarray}

\bibliographystyle{apj}
\bibliography{master}

\end{document}